\documentclass[aps,pra,twocolumn,showpacs,amsmath,amssymb,nofootinbib,superscriptaddress]{revtex4}
\usepackage{amssymb}
\usepackage{stackrel}
\usepackage{amsmath}
\usepackage[english]{babel}
\usepackage{graphicx}
\usepackage{float}
\usepackage{color}

\newcommand{\I}{\mathrm{i}}

\newcommand{\PGR}[1]{{\color{black} #1}}

\begin{document}

\title{A quantum relaxation-time approximation for finite fermion systems}

\author{P.-G.~Reinhard\footnote{Corresponding author : Paul-Gerhard.Reinhard@fau.de}}
\affiliation{Institut f{\"{u}}r Theoretische Physik, Universit{\"{a}}t Erlangen,
             Staudtstra\ss e 7, D-91058 Erlangen, Germany}
\author{E.~Suraud}
\affiliation{Universit\'e de Toulouse; UPS; Laboratoire de Physique
             Th\'{e}orique, IRSAMC; F-31062 Toulouse Cedex, France}
\affiliation{CNRS; UMR5152; F-31062 Toulouse Cedex, France}
\affiliation{Physics Department, University at Buffalo, The State University New York, Buffalo, NY 14260, USA}

\date{\today}

\pacs{05.30.Fk,31.70.Hq,34.10.+x,36.40.Cg}

\begin{abstract}
We propose a relaxation time approximation for the description of the
dynamics of strongly excited fermion systems. Our approach is based on
time-dependent density functional theory at the level of the local
density approximation. This mean-field picture is augmented by
collisional correlations handled in relaxation time approximation
which is inspired from the corresponding semi-classical picture. The
method involves the estimate of microscopic relaxation rates/times
which is presently taken from the well established semi-classical
experience.  The relaxation time approximation implies evaluation of
the instantaneous equilibrium state towards which the dynamical state
is progressively driven at the pace of the microscopic relaxation
time.
As test case, we consider Na clusters of various sizes excited
either by a swift ion projectile or by a short and intense laser pulse,
driven in various dynamical regimes ranging from linear to 
strongly non-linear reactions. We observe a strong effect of
dissipation on sensitive observables such as net ionization and
angular distributions of emitted electrons. The effect is especially
large for moderate excitations where typical relaxation/dissipation
time scales efficiently compete with ionization for dissipating the
available excitation energy. Technical details on the actual procedure
to implement a working recipe of such a quantum relaxation
approximation are given in appendices for completeness.
\end{abstract}
\maketitle

\section{Introduction}

The analysis of the non linear response of finite fermion systems
subject to strong perturbations constitutes a central issue in many
areas of physics. Prominent examples are low energy nuclear physics
and fission \cite{Wad92a,ABe96} as well as excitation of clusters and
molecules by intense laser pulses \cite{Saa06aR,Fen10}. But similar
situations are also encountered elsewhere, for example, in trapped
Fermi gases \cite{Dal99a} or electron transport in nano systems
\cite{Che05aB}.  The case of irradiated clusters and molecules has
become particularly interesting due to recent progress in experimental
techniques at the side of photon sources \cite{Fen10,Kje10} as well as
at the side of detection accessing more and more detailed information
from the decaying system. Particularly useful observables are from the
distributions of emitted electron (energy, angular distributions...)
through elaborate imaging techniques such as Velocity Map Imaging
(VMI) \cite{Wop14,Pin99}.  These highly sophisticated measurements call
for elaborate theoretical modeling to reconstruct the underlying
dynamics of both the irradiation and de-excitation process. The choice
of models ranges from a highly detailed quantum description to
macroscopic rate equations \cite{Fen10}. Each one of these approaches
is valid in a limited range of dynamical scenarios. The robust and
rather versatile Time-Dependent Density Functional Theory (TDDFT),
mostly realized at the level of the Time-Dependent Local-Density
Approximation (TDLDA), \cite{Gro90,Gro96,Mar12} certainly provides one
of the best compromises in the domain of quantum dynamics from the
linear regime up to highly non-linear processes
\cite{Cal00,Mar06,Fen10}.  Still TDLDA basically remains a mean-field
approach and misses by construction dissipative effects from
electron-electron collisions which are expected to play a role in the
course of violent excitation/de-excitation scenarios. This limits the
range of applicability of TDLDA to moderate excitations and/or the
entrance phase of the dynamics. There is thus a strong interest to
extend TDLDA by collisional dynamics in order to push the limits
farther out. It is the goal of this paper to investigate such an
extension by including the effect of collisional correlations (or
dynamical correlations). The notion of ``collisional correlations'' is
taken from Fermi liquid theory \cite{Kad62} with incoherent reduction
of two-body correlations to two-Fermion collisions.  As first example
of application, we discuss the case of irradiated metal clusters.  But
we emphasize that the strategy outlined here is applicable to any
many-fermion system where TDLDA is a good starting point, namely where
it provides a good description of low energy dynamics.

The much celebrated Boltzmann equation constitutes the prototype
approach to collisional dynamics in classical systems \cite{Cer88}.
In quantum systems, one has to account for the Heisenberg uncertainty
relation and the Pauli principle. Pauli blocking can be accounted for
by extending the Boltzmann collision term to the
Boltzmann-Uehling-Uhlenbeck (BUU) form \cite{Ueh33}.  This
semi-classical BUU approach (also known as Vlasov-Uehling-Uhlenbeck
(VUU) equation) provides an acceptable picture at sufficiently large
excitations where quantum shell effects can be ignored. It has been
extensively used in nuclear physics \cite{Ber88,Dur00} and also
explored in metal clusters \cite{Dom98b, Fen04} in a high excitation
domain.  A further limitation of the BUU/VUU approach appears in the
case of clusters: it is not clear that it could be used in systems
others than simple alkalines, where electron wave functions are
sufficiently delocalised and smooth to allow a semi-classical
treatment \cite{Rin80}.  This hinders, e.g., an application to
C$_{60}$ which is one of the systems attracting the most elaborate
analysis of dissipative dynamics so far \cite{Kje10,Han13}.  It should
finally be noted that even in the high-excitation domain a major
de-excitation channel is ionization which may quickly take away large
amounts of excitation energy cooling the system down into a regime
where quantum effects cannot be neglected any more. This limits
BUU/VUU often to the initial stages of a dynamical process. All this
shows that there is an urgent need for a quantum description augmented
by relaxation effects.

In spite of their limitations, semi-classical approaches nevertheless
provide extremely useful guidance in three major aspects. First, they
constitute a long standing testing basis for strategies to implement
approximate relaxation pictures into a quantum framework
\cite{Cal98a}. Second, as we shall exploit below, they serve as source
of inspiration for developing simple approximations for collisional
relaxation in TDLDA. Last, but not least, they deliver approved
estimates for the key quantities (collision rate, intrinsic relaxation
time) to be used in quantum approximations.

The aim of this paper is to propose an extension of (real-time) TDLDA
by collisional correlations. We shall  exploit the  experience from
semi-classical approximations but keep the level of description
quantum mechanical throughout.
The paper is organized as follows. Section \ref{sec:mf} provides the
basics of the mean field description and introduces associated
notations. Section \ref{sec:RTA} describes the proposed
scheme for the Relaxation Time Approximation (RTA) and details of its
handling. First results for the case of metal clusters are then
presented in section \ref{sec:res} and conclusions and perspectives
are drawn in section \ref{sec:conc}. Finally, the text is completed by
a series of appendices which give some technical details on the RTA
and its handling in practice.

\section{Mean-field propagation}
\label{sec:mf}

The starting point and dominant feature of the dynamics is the
propagation at the level of the mean field. In this paper, we are
dealing with the electron dynamics in metal clusters and we describe
it by time-dependent density functional theory at the level of the
Time-Dependent Local-Density Approximation (TDLDA) treated in the real
time domain \cite{Gro90,Gro96}.  It is augmented by a self-interaction
correction (SIC) approximated by average-density SIC (ADSIC)
\cite{Leg02} in order to attain correct ionization properties
\cite{Klu13} in the course of the dynamical simulation. TDLDA is
formulated within the usual Kohn-Sham picture in terms of a set of
occupied single-particle (s.p.) wavefunctions
$\{|\phi_\alpha\rangle,\alpha=1...N\}$. Their dynamics is described by
the time-dependent Kohn-Sham equation
\begin{equation}
  \I\partial_t|\phi_\alpha\rangle
  =
  \hat{h}[\varrho]|\phi_\alpha\rangle
\label{eq:KSwf}
\end{equation}
where $\hat{h}$ is the Kohn-Sham mean-field Hamiltonian which is a
functional of the instantaneous local density
$\varrho(\mathbf{r},t)=\sum_\alpha|\phi_\alpha(\mathbf{r},t)|^2$
\cite{Dre90,Rei04aB}. The time evolution delivered by
Eq. (\ref{eq:KSwf}) can be expressed formally by the
unitary one-body  time-evolution operator 
\begin{subequations}
\begin{equation}
  \hat{U}(t,t')
  =
  \hat{\mathcal{T}}\mathrm{ exp}\left(-i \int_t^{t'} \hat{h}(t'')dt''\right)
\label{eq:KSpropag}
\end{equation}
where $\hat{\mathcal{T}}$ is the time-ordering operator.
This yields a closed expression for the time-evolution of s.p. states
\begin{equation}
  |\phi_\alpha(t)\rangle
  =
  \hat{U}(t,t')|\phi_\alpha(t')\rangle.
\label{eq:KSpropag2}
\end{equation}
\end{subequations}

So far, TDLDA propagates pure states. Dissipation which we will add
later on leads inevitably to mixed states. This requires to generalize
the description from fully occupied s.p. wavefunctions to a one-body
density operator $\hat{\rho}$. It is often denoted as one-body density
matrix $\rho(\mathbf{r},\mathbf{r}')$ which is, in fact, the
coordinate representation of $\hat{\rho}$. A (natural orbitals)
representation of the one-body density operator in terms of
s.p. wavefunctions reads
\begin{equation}
  \hat{\rho}
  =
  \sum_{\alpha=1}^\infty|\phi_\alpha\rangle W_\alpha\langle\phi_\alpha|
  \quad.
\label{eq:rhodiag}
\end{equation}
New are here the weights $W_\alpha$, the probability with which a
state $|\phi_\alpha\rangle$ is occupied. The mean-field propagation
(\ref{eq:KSwf}) then becomes
\begin{equation}
  \I\partial_t\hat{\rho}
  =
  \left[\hat{h}[\varrho],\hat{\rho}\right]
\label{eq:KSrho}
\end{equation}
where $\hat{h}[\varrho]$ is formally the same as before and the
local density is now computed as
\begin{equation}
\varrho(\mathbf{r},t)=\sum_\alpha{W}_\alpha|\phi_\alpha(\mathbf{r},t)|^2 .
\label{eq:rhoW}
\end{equation}
The pure mean-field propagation (\ref{eq:KSrho}) leaves the occupation
weights $W_\alpha$ unchanged and propagates only the s.p. states.  The
mean-field propagation of an initial state (\ref{eq:rhodiag}) then
reads
\begin{eqnarray}
  \hat{\rho}(t)
  &=&
  \sum_{\alpha=1}^\infty
  |\phi_\alpha(t)\rangle W_\alpha\langle\phi_\alpha(t)|
\nonumber\\  
  &=&
  \hat{U}(t,0)\hat{\rho}(0)\hat{U}^{-1}(t,0)
\label{eq:KSrhoevol}
\end{eqnarray}
where $\hat{U}$ is the mean-field evolution operator (\ref{eq:KSpropag}).

\section{The relaxation-time approximation (RTA) for finite fermion systems}
\label{sec:RTA}

\subsection{Motivation: RTA in a semi-classical framework}

In homogeneous fermion systems the phase space distribution $f$ only
depends on momentum and dynamical correlations can be described by the
Uehling-Uhlenbeck collision term $I_\mathrm{UU}[f(\mathbf{p})]$
\cite{Ueh33,Pin66}. It is a functional of the momentum space
distribution $f(\mathbf{p})$ which drives the dynamics steadily
towards the thermal equilibrium distribution $f_\mathrm{eq}$. As the
collision term $I_\mathrm{UU}$ conserves particle number, mean
momentum and energy, the
$f_\mathrm{eq}(\mathbf{p};\varrho,\mathbf{j},E)$ represents the
equilibrium for given density $\varrho$, current $\mathbf{j}$, and
kinetic energy $E$. Sufficiently close to equilibrium, one
can approximate the convergence as exponential relaxation which allows
to model the dynamical process simply as
\begin{equation}
  \partial_tf(\mathbf{p},t)
  =
  -\tau_\mathrm{relax}^{-1}
  \left(f(\mathbf{p},t)-f_\mathrm{eq}(\mathbf{p};\varrho,\mathbf{j},E)\right)
\label{eq:RTAclass}
\end{equation}
where $\tau_\mathrm{relax}$ is the relaxation time. The expectation
values $\varrho,\mathbf{j},E$ for density, current, and energy are the
ones associated to $f(\mathbf{p},t)$ and computed as 
$\varrho=\int d^3p\,f(\mathbf{p},t)$, 
$\mathbf{j}=\int d^3p\,\mathbf{p}\,f(\mathbf{p},t)$, 
and $E=\int d^3p\,\mathbf{p}^2\,f(\mathbf{p},t)/(2m)$.  
This is called the Relaxation Time Approximation (RTA). It was
introduced in \cite{Bha54} and it has been used used in that form for
a wide variety of homogeneous systems \cite{Pin66,Ash76}.

Finite fermion systems are spatially inhomogeneous.  An obvious
generalization to this case is to extend $f(\mathbf{p})$ to a
phase-space distribution $f(\mathbf{r},\mathbf{p})$. This leads to a
semi-classical description which is valid for high excitations (or
temperatures) where quantum shell effects are obsolete. The
mean-field dynamics is then described by the Vlasov equation and the
dynamical correlations by an additional collision term
$I_\mathrm{UU}[f(\mathbf{r},\mathbf{p})]$ yielding together the
Vlasov-Uehling-Uhlenbeck (VUU) equation \cite{Lif88}
\begin{equation}
  \partial_t f-\left\{h,f\right\}
  =
  I_\mathrm{UU}[f(\mathbf{r},\mathbf{p})]
\label{eq:VUU}
\end{equation}
where $h$ is the (classical) mean field Hamiltonian.  
In this semi-classical approach,
collisions
are local,  changing for a given
$\mathbf{r}$ only the momentum distribution at this point. It thus
  establishes local conservation laws \cite{Gue88a} such that
  collisional relaxation conserves local density
$\varrho(\mathbf{r},t)$, local current $\mathbf{j}(\mathbf{r},t)$, and
local kinetic energy $E_\mathrm{kin}(\mathbf{r},t)$.  The
collision term thus drives towards a local and instantaneous
equilibrium
$f_\mathrm{eq}(\mathbf{r},\mathbf{p};\varrho,\mathbf{j},E_\mathrm{kin})$.
The global equilibration is achieved at slower pace by
interplay with the long range transport described by the mean-field
propagation (Vlasov part of the VUU equation). The RTA for the VUU
equation (\ref{eq:VUU}) reads \cite{Bha54,Koh80a,Koh82a,Won83a}
\begin{equation}
  \partial_t f-\left\{h,f\right\}
  =
  -\frac{1}{\tau_\mathrm{relax}}
  \left(f(\mathbf{r},\mathbf{p},t)
    -f_\mathrm{eq}(\mathbf{r},\mathbf{p};\varrho,\mathbf{j},E_\mathrm{kin})\right)
\label{eq:VUUrelax}
\end{equation}
where the constraints \PGR{$\varrho,\mathbf{j},E_\mathrm{kin}$} depend
on position $\mathbf{r}$ and time $t$. This is the model which we will
now generalize to the case of quantum mean-field theory.

\subsection{RTA in quantum-mechanical framework}

The generalization of the one-body phase-space distribution
$f(\mathbf{r},\mathbf{p})$ to a quantum-mechanical mean-field theory
is the one-body density operator $\hat{\rho}$, or one-body density
matrix $\rho(\mathbf{r},\mathbf{r}')$ respectively. The equation of
motion for $\hat{\rho}$ including dynamical correlations reads in
general \cite{Rei85f,Goe86a}
\begin{eqnarray}
  \mathrm{i}\partial_t\hat{\rho}
  -
  \big[\hat{h},\hat{\rho}\big]
  &=&
  \hat{I}[\hat{\rho}]
  \quad.
\label{eq:EoMfull}
\end{eqnarray}
The left hand side embraces the mean-field propagation. It may be
time-dependent Hartree-Fock or the widely used LDA version of TDDFT.
The right-hand side consists in the quantum-mechanical collision term.
Motivated by the successful semi-classical RTA, we import
Eq.~(\ref{eq:VUUrelax}) for the quantum case as
\begin{eqnarray}
  \partial_t\hat{\rho}
  &=&
  -\mathrm{i}\big[\hat{h},\hat{\rho}\big]
  -
  \frac{1}{\tau_\mathrm{relax}}
  \left(\hat{\rho}-\hat{\rho}_\mathrm{eq}[\varrho,\mathbf{j},E]\right)
  \;,
\label{eq:EoMbasic}
\end{eqnarray}
where $\hat{\rho}_\mathrm{eq}$ is the density operator of the thermal
equilibrium for local density $\varrho(\mathbf{r},t)$, current
distribution $\mathbf{j}(\mathbf{r},t)$ and total energy $E(t)$
given at that instant of time $t$.  These
constraining conditions are, in fact, functionals of the actual state
$\hat{\rho}$, i.e. $\varrho[\hat{\rho}]$, $\mathbf{j}[\hat{\rho}]$,
and $E[\hat{\rho}]$.  For the diagonal representation
Eq.(\ref{eq:rhodiag}) of the density operator $\hat{\rho}$, they read
\begin{subequations}
\begin{eqnarray}
  \varrho(\mathbf{r})
  &=&
  \sum_\alpha \left|\phi_\alpha(\mathbf{r})\right|^2 W_\alpha
  \quad,
\\
  \mathbf{j}(\mathbf{r})
  &=&
  \sum_\alpha W_\alpha\phi_\alpha^*(\mathbf{r})
     \frac{\stackrel{\rightarrow}{\nabla}-\stackrel{\leftarrow}{\nabla}}
          {2\mathrm{i}}
     \phi_\alpha(\mathbf{r})
  \quad.
\end{eqnarray}
\end{subequations}
The energy $E(t)$ is taken as the total energy because the
semi-classical concept of a \PGR{local kinetic energy} is
ambiguous in a quantum system.  This RTA equation (\ref{eq:EoMbasic})
looks innocent, but is very involved because many entries depend in
various ways on the actual state $\hat{\rho}(t)$. The self-consistent
mean field is a functional of the actual local density,
i.e. $\hat{h}=\hat{h}[\varrho]$. The instantaneous equilibrium density
$\hat{\rho}_\mathrm{eq}$ is the solution of the stationary, thermal
mean-field equations with constraint on the actual
$\varrho(\mathbf{r})$, $\mathbf{j}(\mathbf{r})$ and energy $E$, for
details see Appendix \ref{sec:hdenscurrE}.

The relaxation time $\tau_\mathrm{relax}$ is estimated in
semi-classical Fermi liquid theory, for details see appendix
\ref{sec:relaxtime}. For the metal clusters serving as test examples
in the following, it becomes
\begin{equation}
  \frac{\hbar}{\tau_\mathrm{relax}}
  =
  {0.40}\frac{\sigma_{ee}}{r_s^2}\frac{{E}^*_\mathrm{intr}}{N}
  \quad,
\label{eq:relaxtime}
\end{equation}
where $E^*_\mathrm{intr}$ is the intrinsic (thermal) energy of the
system (appendix \ref{app:eintr}), $N$ the actual number of particles,
$\sigma_{ee}$ the in-medium electron-electron cross section, and
$r_s$ the effective Wigner-Seitz radius of the electron cloud.

\subsection{Summary of the procedure}
\label{sec:summary}

\begin{figure*}
\setlength\unitlength{1mm}
\thicklines
\begin{center}
\fbox{
\begin{picture}(151,101)(0,103)
\put(0,200){\mbox{starting point: 
  $\hat{\rho}(t)
  =
  \sum_\alpha|\phi_\alpha(t)\rangle W_\alpha(t)\langle\phi_\alpha(t)|$
}}
\put(0.8,190.2){\mbox{1}}
\put(1.42,191){\circle{3.5}}
\put(5,198){\vector(0,-1){14}}
\put(7,190){\mbox{mean-field propagation: 
$\begin{array}{l}
 |\phi_\alpha^\mathrm{(mf)}\rangle=\hat{U}(t+\Delta t,t)|\phi_\alpha(t)\rangle
\\
 W_\alpha(t)=W_\alpha^{(mf)}=\mbox{const.}
\end{array}$
}}
\put(0,180){\mbox{
  $ \hat{\rho}_{mf} =\hat{\rho}_{mf}(t + \Delta t) 
  =
  \sum_\alpha|\phi_\alpha^{(mf)}\rangle W_\alpha^{(mf)}\langle\phi_\alpha^{(mf)}|$
}}
\put(70,181){\vector(1,0){20}}
\put(90,180){\mbox{  $\varrho_{mf}(\mathbf{r},t+\Delta t)\,,\,\mathbf{j}_{mf}(\mathbf{r},t+\Delta t),E_{mf}$}}
\put(79.75,177.6){\mbox{2}}
\put(80.42,178.5){\circle{3.5}}

\put(104.8,169.5){\mbox{3}}
\put(105.42,170.5){\circle{3.5}}
\put(110,175){\vector(0,-1){10}}
\put(92,152){\mbox{
\begin{minipage}{8cm}
\begin{flushleft}
  density-constrained mean field (DCMF) 
\\[3pt]
1. 
$\begin{array}[t]{rcl}
  \hat{\rho}_{eq}
  &=&
  \hat{\rho}_{eq}[\varrho_{mf}(\mathbf{r})\,,\,\mathbf{j}_{mf}(\mathbf{r}),E_{mf}]
\\[3pt]
  &=&
  \sum_\alpha|\phi'_\alpha\rangle  W'_\alpha\langle\phi'_\alpha|
\end{array}$
\\[4pt]
2. intrinsic excitation energy $E^*_\mathrm{intr}$
\end{flushleft}
\end{minipage}
}}
\put(138,143){\vector(-1,-2){7}}
\put(95,129){\mbox{
\begin{minipage}{8cm}
\begin{flushleft}
relaxation time:
\\[4pt]
$\hbar\tau_\mathrm{relax}^{-1}=0.40\,\sigma_{ee}r_s^{-2}\,E^*_\mathrm{intr}/N$
\end{flushleft}
\end{minipage}
}}
\put(4,177){\vector(0,-1){39}}
\put(94,153.5){\line(-3,-1){45}}
\put(54.7,142.5){\mbox{4}}
\put(55.6,143.5){\circle{3.5}}
\put(53.7,128.5){\mbox{4}}
\put(54.6,129.5){\circle{3.5}}
\put(0,154.5){\mbox{4}}
\put(0.9,155.5){\circle{3.5}}
\put(49,138.7){\vector(0,-1){3.7}}
\put(0,119.5){\mbox{
\begin{minipage}{8cm}
\begin{flushleft}
$\displaystyle
\hat{\rho}(t\!+\!\Delta t) 
= 
\hat{\rho}_{mf} -
\frac{\Delta t}{\tau_\mathrm{relax}}\left[\hat{\rho}_{mf}-\hat{\rho}_{eq}\right]
$
\\[12pt]
\hspace*{1.8em}diagonalize to natural orbitals:
\\[4pt]
  $\hat{\rho}(t\!+\!\Delta t)
  =
  \sum_\alpha|\phi_\alpha(t\!+\!\Delta t)\rangle \tilde{W}_\alpha
  \langle\phi_\alpha(t\!+\!\Delta t)|$
\\[12pt]
\hspace*{1.8em}final fine-tuning of $W_\alpha$ to reproduce $E_{mf}$
\\[4pt]
  $\hat{\rho}(t\!+\!\Delta t)
  =
  \sum_\alpha|\phi_\alpha(t\!+\!\Delta t)\rangle W_\alpha(t\!+\!\Delta t)
  \langle\phi_\alpha(t\!+\!\Delta t)|$
\end{flushleft}
\end{minipage}
}}
\put(94,127.5){\line(-1,0){53}}
\put(41,127.5){\vector(-2,1){6}}
\put(4,130.5){\vector(0,-1){9}}
\put(0,126.5){\mbox{5}}
\put(0.9,127.5){\circle{3.5}}
\put(4,116.5){\vector(0,-1){8.5}}
\put(0,112.5){\mbox{6}}
\put(0.9,113.5){\circle{3.5}}
\end{picture}
}
\end{center}
\caption{\label{fig:summary} Sketch of the scheme for performing one
  large time step $t\longrightarrow t\!+\!\Delta t$ in solving the RTA
  equations.  The numbers in open circles indicate the steps as
  outlined in the text.  }
\end{figure*}
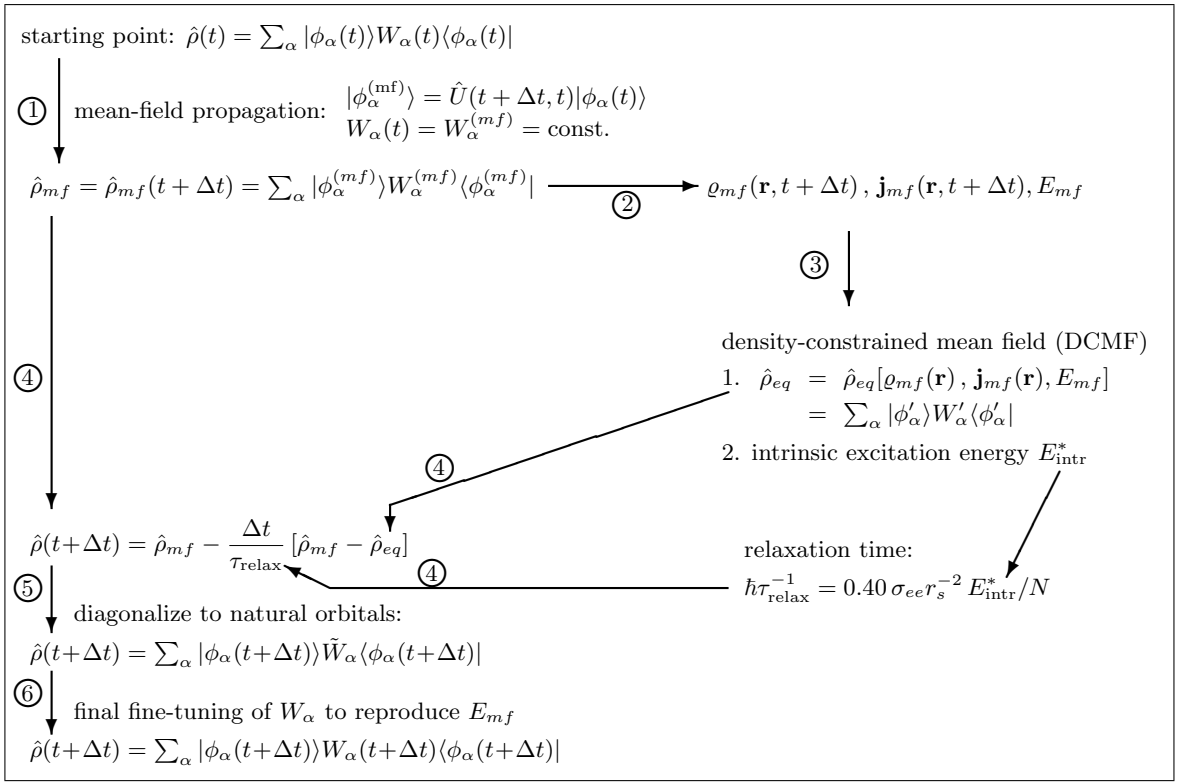

The solution of the RTA equations is rather involved. We explain the
necessary steps here from a practical side and unfold details in the
appendices. We briefly summarize the actual scheme for one step from
$t$ to $t\!+\!\Delta t$.  Note that mean-field propagation (actually
TDLDA) runs at a much faster pace than relaxation.  We resolve it by
standard techniques \cite{Cal00,Rei04aB} on a time step $\delta t$
which is much smaller (factor 10--100) than the RTA step $\Delta
t$. We summarize this TDLDA propagation in the evolution operator
$\hat{U}$ from Eq.~(\ref{eq:KSpropag}) and discuss only one RTA
step. Its sub-steps are sketched in Figure \ref{fig:summary} and
explained in the following whereby the label here correspond to the
ones in the Figure:
\begin{enumerate}
   \item\label{it:TDLDA} We first propagate $\hat{\rho}$ by pure
     TDLDA.  This means that the s.p. states in representation
     (\ref{eq:rhodiag}) evolve as
     $|\phi_\alpha(t)\rangle\rightarrow
     |\phi_\alpha^\mathrm{(mf)}\rangle=\hat{U}(t+\Delta
     t,t)|\phi_\alpha(t)\rangle$, while the occupation weights
     $W_\alpha$ are kept frozen (pure mean-field propagation).
   \item\label{it:newrho} We compute density
     $\varrho(\mathbf{r},t+\Delta t)$, current
     $\mathbf{j}(\mathbf{r},t+\Delta t)$, and total energy
     $E_\mathrm{mf}$ associated to the TDLDA-propagated density matrix
     $\hat{\rho}_\mathrm{mf}$.
   \item\label{it:DCMF} We determine the thermal mean-field
     equilibrium state $\hat{\rho}_\mathrm{eq}$ constrained to the
     given $\varrho$, $\mathbf{j}$, and $E_\mathrm{mf}$.  This is
     achieved by Density-Constrained Mean Field (DCMF) iterations 
     as outlined in
     Appendix \ref{sec:hdenscurrE}.  The actual equilibrium state
     $\hat{\rho}_\mathrm{eq}$ is represented by new s.p. states
     $\{|\phi'_{\alpha}\rangle\}$ and new occupation numbers $W'_\alpha$
     in diagonal form (\ref{eq:rhodiag}).
   \item \label{it:compo} We compose the new density matrix from the
     TDLDA propagated state $\hat{\rho}_\mathrm{mf}$ and the
     equilibration driving term
     $\hat{\rho}_\mathrm{mf}-\hat{\rho}_\mathrm{eq}$ with the
     appropriate weight $\Delta t/\tau_\mathrm{relax}$, as outlined in
     Appendix \ref{sec:mix}.  The relaxation time
     Eq. (\ref{eq:relaxtime}) requires the actual intrinsic excitation
     energy $E^*_\mathrm{intr}$ which is also obtained from DCMF, see
     appendix \ref{app:eintr}.
   \item \label{it:natural} 
     We diagonalize the state emerging from
     step \ref{it:compo} to natural-orbital representation
     Eq. (\ref{eq:rhodiag}).  This yields the s.p. states
     $\{|\phi_\alpha(t\!+\!\Delta{t})\rangle\}$ for the next step and
     preliminary new occupations $\tilde{W}_\alpha$.
   \item \label{it:therm} 
     After all these steps, the initial energy
     $E_\mathrm{mf}=E_\mathrm{TDLDA}(t)$ may not be exactly
     reproduced. We may remain with a small energy mismatch as
     compared to the goal $E_\mathrm{mf}$.  We now apply a small
     iterative thermalization step to readjust the energy, as outlined
     in Appendix \ref{sec:corriter}. This then yields the final
     occupation weights $W_\alpha(t\!+\!\Delta{t})$ which comply with
     energy conservation.
\end{enumerate}
The scheme can be used also in connection with absorbing boundary
conditions \cite{Cal00,Rei06c}. The particle loss will be mapped
automatically to loss of occupation weights in step \ref{it:compo}. A
word is in order about the choice of the time steps. The $\delta t$ for
propagation of TDLDA is limited by the maximal energy on the grid
representation and thus very small (for Na clusters typically 0.005
fs). The stepping for the relaxation term needs only to resolve the
changes in the actual mean field which is achieved already with
$\Delta t\approx 0.5$ fs. We have tested a sequence of $\Delta t$ and
find the same results for all $\Delta t\leq 0.5$ fs. Changes appear
slowly above that value.  For reasons of efficiency, we thus use the
largest safe value of $\Delta t= 0.5$ fs.

\subsection{Numerical representation and computation of relevant observables  }
\label{sec:observ}

The numerical implementation of TDLDA is done in standard
manner~\cite{Cal00,Rei04aB}.  The coupling to the ions is mediated by
soft local pseudopotentials~\cite{Kue99}.  The Kohn-Sham potential is
handled in the Cylindrically Averaged Pseudo-potential Scheme (CAPS)
\cite{Mon94a,Mon95a}, which has proven to be an efficient and reliable
approximation for metal clusters close to axial symmetry.
Wavefunctions and fields are thus represented on a 2D cylindrical grid
in coordinate space \cite{Dav81a}.  For the typical example of the
Na$_{40}$ cluster, the numerical box extends up to 104 a$_0$ in radial
direction and 208 a$_0$ along the $z$-axis, while the grid spacing is
0.8 a$_0$. To solve the (time-dependent) Kohn-Sham equations
(\ref{eq:KSwf}) we use time-splitting for time
propagation~\cite{Fei82} and accelerated gradient iterations for the
stationary solution \cite{Blu92}. The Coulomb field is computed with
successive over-relaxation \cite{Dav81a}.  We use absorbing boundary
conditions~\cite{Cal00,Rei06c}, which gently absorb all outgoing
electron flow reaching the bounds of the grid and thus prevent
artifacts from reflection back into the reaction zone.  We take the
exchange-correlation energy functional from Perdew and
Wang~\cite{Per92}.

A great manifold of observables can be deduced from the
$\hat{\rho}(t)$ thus obtained. We will consider in the following the
dipole signal, dipole spectrum, ionization, angular distribution of
emitted electrons, and entropy. We focus here on the dipole moment
along symmetry axis $z$, which is obtained from the local density as
$\langle\hat{d}_z\rangle(t)=\int{d}^3r\,d_z(\mathrm{z})\varrho(\mathrm{r})$
where $d_z(\mathrm{z})=z$ is the (local) dipole operator. The dipole
strength distribution is computed with the methods of spectral
analysis \cite{Cal97b}. It is attained by an instantaneous
dipole-boost excitation, collecting $\langle\hat{d}_z\rangle(t)$
during propagation, and finally Fourier transforming
$\langle\hat{d}_z\rangle(t)$ into frequency domain. The angular
  distribution of emitted electrons is obtained from recording the
absorbed electrons as in TDLDA \cite{Poh04b,Rei06aR}. The angular
distribution is characterized by the anisotropy parameter
$\beta_2$, the leading parameter in the photo-electron angular cross
section $d\sigma/d\Omega \propto (1+\beta_2 P_2(cos(\theta)+....)$
\cite{Wop10a,Wop10b} where $P_2$ is the second order Legendre
polynomial and $\theta$ the direction with respect to laser
polarization axis (here $z$-axis in 2D cylindrical geometry). A
specific quantity to track relaxation processes is the one-body
entropy which is computed in diagonal representation
(\ref{eq:rhodiag}) by the standard expression \cite{Rei98aB}
\begin{equation}
  S
  =
  - \sum_\alpha\left[
    W_\alpha\log W_\alpha
    +
    (1\!-\!W_\alpha)\log (1\!-\!W_\alpha)
  \right]
\label{eq:entropy}
\end{equation}
in units of Bolzmann constant. 
It serves as a direct indicator of thermalization and allows to 
read off the typical time scale of relaxation processes.

\section{Results}
\label{sec:res}

\subsection{The test cases}

As test cases, we will consider the clusters, Na$_{40}$,
Na$_{9}^+$, Na$_{41}^+$, and Na$_{93}^+$. All test cases have
electronic shell closures ($N_\mathrm{el}=8,40,92$ \cite{Bra93}) and
are thus close to spherical symmetry. This is no principle restriction
because computations with deformed systems show
similar pattern. In fact, shell closures with their large HOMO-LUMO
gaps are the most demanding situations (thus critical test cases) for
the RTA scheme. The ionic ground-state configuration is optimized by
iterative cooling in the spirit of simulated annealing
\cite{Cal00,Rei04aB}. In the following we are interested exclusively
in electronic dissipation and we are considering rather short time
intervals. We thus keep the ions frozen at their ground-state
configurations. The Wigner-Seitz radius required in the estimate of
local relaxation time Eq. (\ref{eq:relaxtime}) is computed with the
recipes of Appendix \ref{sec:relaxtime}. It turns out to be almost the
same for all test cases mentioned above. We use in practice $r_s=3.7$
a$_0$. The TDLDA equations are propagated with a time step of $\delta
t = 0.005$ fs. The larger time step for evaluation of the dissipative
term is $\Delta t=0.5$ fs.

\subsection{Trends for boost excitations}

In the first round, we use the simplest excitation mechanism to
elucidate the basic effects of dissipative term. This is an
instantaneous dipole boost
$\phi_\alpha\rightarrow\exp(-\I\,p_0\,\hat{d}_z)\phi_\alpha$ applied
to all s.p. wavefunctions in the same manner \cite{Cal00,Rei04aB}. The
boost momentum $p_0$ regulates its strength. We are characterizing the
boost strength henceforth in terms of the initial excitation energy
$E^*_0=Np_0^2/(2m)$ brought into the cluster by the boost. Mind that
$E^*_0$ is the {\it initial} excitation energy deposited into the
system, not to be confused with the time dependent intrinsic
excitation energy $E^*_\mathrm{intr}$ used in estimating the
relaxation time (\ref{eq:relaxtime}). The instantaneous dipole boost
models to a good approximation the time-dependent Coulomb field at the
cluster site for collisions with very fast ion passing by the cluster
\cite{Bae06a,Wop14aR}.

\subsubsection{Optical response as initial example}
\label{sec:optresp}

Optical response is the basic observable characterizing the reaction
of a cluster to an electromagnetic perturbation and it serves as key
to the analysis of a large variety of dynamical scenarios
\cite{Rei04aB}. It is thus of interest to check the impact of RTA on
optical response. We take here Na$_{40}$ as test example. Its optical
response is dominated by a surface plasmon resonance in a rather
narrow spectral range around 2.7 eV.

\begin{figure}
\centerline{\includegraphics[width=0.99\linewidth]{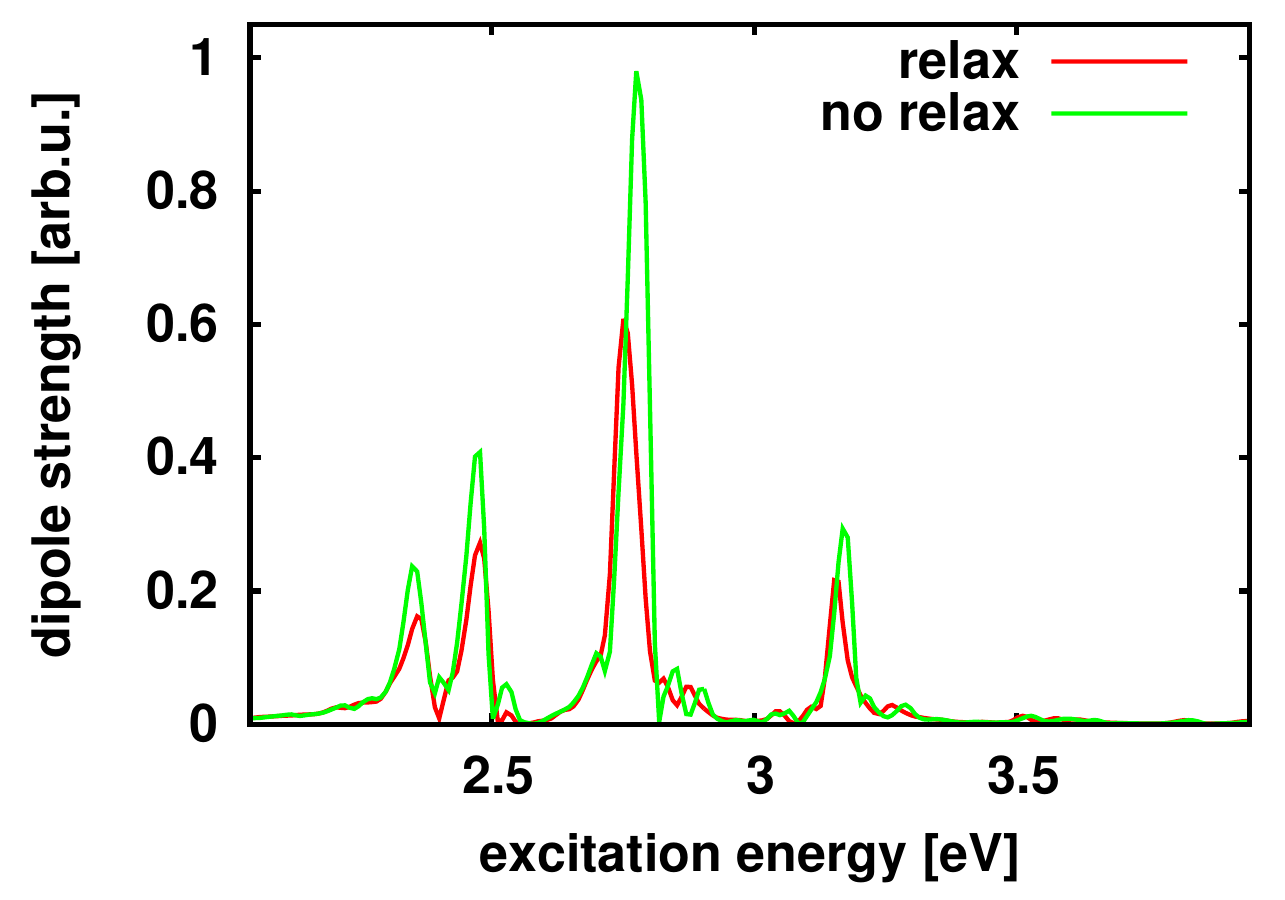}}
\caption{\label{fig:Na40ion-varyboost13-dipstr} 
 (Color online) Spectral distribution
  of dipole strength for Na$_{40}$ with ionic background in the CAPS,
  evaluated for a boost strength $E^*_0=2.7$ eV according to the scheme of
  \cite{Cal97b}. Compared are calculations with and without
  relaxation term.
  }
\end{figure}
Figure \ref{fig:Na40ion-varyboost13-dipstr} shows the effect of
dissipation on the spectral distribution of dipole strength for an
excitation in the one-plasmon regime, i.e. $E^*_0=2.7$ eV. The spectra
are computed after instantaneous boost with spectral analysis of the
dipole signal \cite{Cal97b}. As expected, the relaxation term leads to
broadening of the spectral peaks because the lifetime of the
eigenmodes is reduced by dissipation. However, the effect is
surprisingly small. This is due to the competition with an even
stronger relaxation through direct electronic emission.  This
mechanism is contained in TDLDA and has already achieved a great deal
of smoothing the peaks leaving little to do for collisional
relaxation.  But mind that this competition between  direct
emission and collisional relaxation depends sensitively on the actual
details of the system and excitation mechanisms, especially on the
total amount of deposited excitation energy $E_0^*$.

\subsubsection{Trends with varied excitation strength}

\begin{figure}
\centerline{\includegraphics[width=0.9\linewidth]{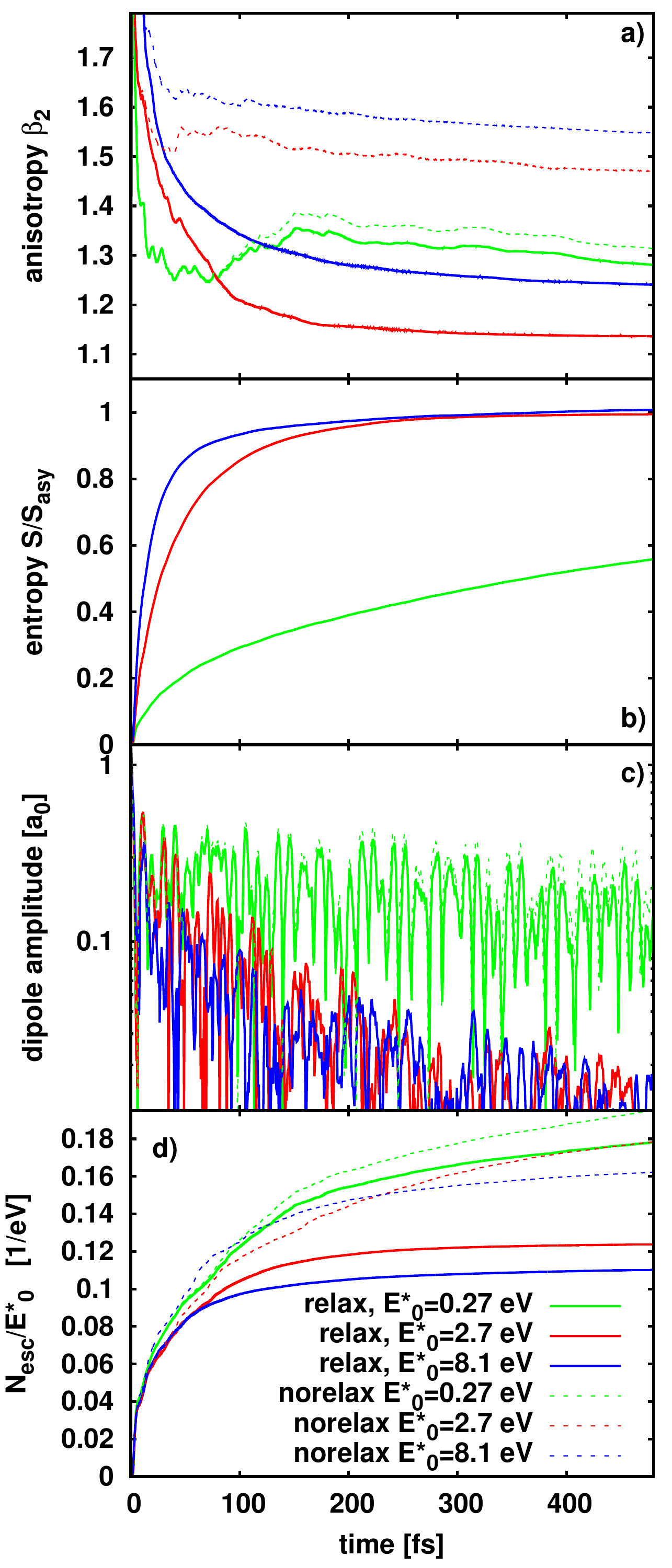}}
\caption{\label{fig:Na40ion-varyboost13} 
 (Color online) Time evolution of four key
  observables after an instantaneous boost with three different boost
  energies as indicated. Test case is Na$_{40}$ with ionic background
  in the CAPS. The observables are: d) ionization $N_\mathrm{esc}/E^*_0$
  rescaled to initial excitation energy $E^*_0$, c) envelope of the
  dipole signal $\langle\hat{d}_z\rangle$, b) entropy $S$ relative to the
  asymptotic value $S_\mathrm{asy}$, and a) anisotropy $\beta_2$ of
  the angular distribution of emitted electrons.  The cases with
  relaxation are drawn with full lines and
  computed with the standard scattering cross section $\sigma_{ee}=6.5$
  a$_0^2$. TDLDA results are drawn with dotted lines.}
\end{figure}
Figure \ref{fig:Na40ion-varyboost13} shows the time evolution of
relative ionization $N_\mathrm{esc}/E^*_0$, envelope of the dipole
signal $\langle\hat{d}_z\rangle$, entropy $S$, and anisotropy
$\beta_2$ for Na$_{40}$ after boosts with different strengths. Results
from pure TDLDA are shown with dotted lines and those including
dissipation with full lines. The three cases for $E^*_0$ represent
three regimes: linear response for $E^*_0=0.27$ eV, the one-plasmon
regime for $E^*_0=2.7$ eV, and the three-plasmon (non-linear) regime
for $E^*_0=8.1$ eV.

The dipole amplitudes (panel c) are shown in logarithmic scale to
visualize the long-time trend. They decay with time, the faster the
higher the excitation $E^*_0$.  Even the envelope of the dipole
signal is heavily fluctuating such that it is hard to read off a
global relaxation time. Moreover, already the TDLDA amplitudes are
attenuated due to Landau damping and direct electron emission
\cite{Cal00,Rei04aB}.  The dissipative effect corresponds to the
difference between TDLDA and RTA propagation. Unfortunately, it cannot
be extracted cleanly from the dipole signal.

The entropy, Eq. (\ref{eq:entropy}), stays at $S=0$ for pure TDLDA.
It is thus a selective signal for dissipative effects and only results
from the RTA calculations are relevant here.  Panel b in Figure
\ref{fig:Na40ion-varyboost13} shows the entropy relative to the
asymptotic value to allow a direct comparison of the three cases.  All
three cases show a nice exponential convergence. The corresponding
global relaxation time depends sensitively on the excitation
energy $E^*_0$. There is practically no dissipation for the faintest
excitation. This is clear from the recipe Eq. (\ref{eq:relaxtime}) for
the local relaxation rate which is $\propto E^*_0$ and thus disappears
for $E^*_0\rightarrow 0$. Dissipation increases with increasing
excitation amounting to a global relaxation time of about 50 fs for
the one-plasmon regime $E^*_0=2.7$ eV and to 25 fs for $E^*_0=8.1$ eV.

Dissipation has a large effect on ionization (panel d). For TDLDA, the
leading damping mechanism in the late phase is electron emission and
this continues for long as one can see from the slowly but
continuously growing $N_\mathrm{esc}$, accompanied by slowly
decreasing dipole amplitude. Dissipation offers an alternative channel
for damping, namely internal excitation. Consequently, ionization is
much suppressed, the more the stronger dissipation is at work,
i.e. practically not for $E^*_0= 0.27$ eV and increasing with $E^*_0$.

Anisotropy is a well known signal of thermalization. Angular
distributions become more and more isotropic the more thermalized a
system is. It is thus important to compare TDLDA with RTA in that
respect. This is done in panel a of Figure
\ref{fig:Na40ion-varyboost13}.  Dissipation clearly leads to a
reduction of $\beta_2$. It is, nevertheless, interesting to note the
counteracting effects pulling on $\beta_2$, one due to TDLDA, the
other one to RTA. The trend is simple for pure TDLDA: the stronger the
boost the larger the drive to forward/backward emission and thus the
larger $\beta_2$.  On the other hand, larger excitation enhances
dissipation and reduces increasingly $\beta_2$. Eventually, the trend
can go both ways.

Finally, remind that the analysis of angular distributions is done
here over finite times which means that the $\beta_2$ shown here are
not yet the asymptotic values. The excitation energy stored in
intrinsic degrees of freedom by dissipation will be released slowly
later, to a large extent in terms of thermal electron emission. This
adds an isotropic background of thermal electrons which will further
reduce the anisotropy. A strategy to estimate this effect is proposed
in \cite{Wop14aR,Wop14b}. We skip this final clean-up in the present
exploratory study as it is not crucial for understanding RTA.

\subsubsection{Trends with electron-electron cross section}

\begin{figure}
\centerline{\includegraphics[width=0.9\linewidth]{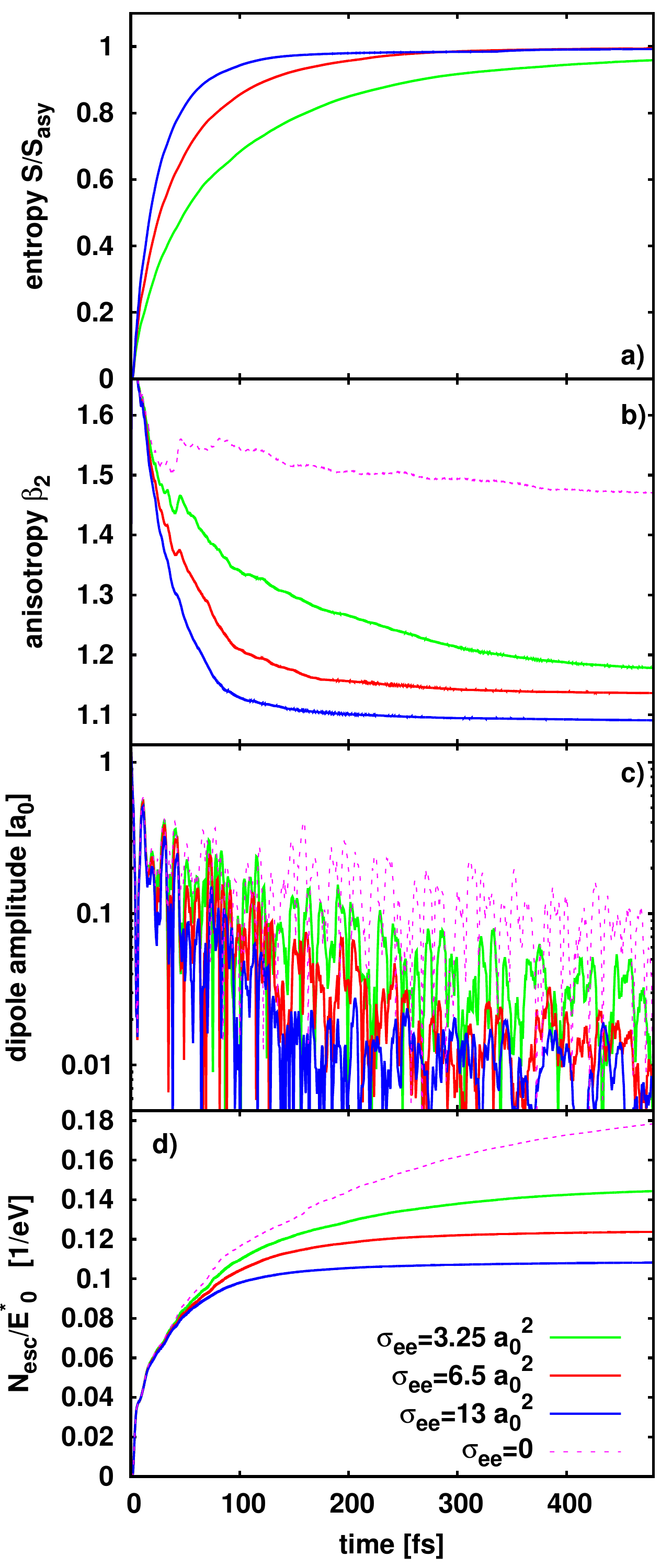}}
\caption{\label{fig:Na40ion-varystrength} 
 (Color online) Same as Fig.~\ref{fig:Na40ion-varyboost13}, but for varied 
   cross sections $\sigma_{ee}$ 
  defining $\tau_\mathrm{relax}$ which correspond to the
  standard  $\sigma_{ee}=6.5$ a$_0^2$ and half as well
  as twice that value. The  boost energy is $E^*_0=2.7$ in all cases.}
\end{figure}
The choice of the cross section $\sigma_{ee}$ for in-medium electron
scattering is, of course, a crucial ingredient for the estimate of the
relaxation time Eq. (\ref{eq:relaxtime}).  To explore the sensitivity
on the choice of $\sigma_{ee}$, we show in Figure
\ref{fig:Na40ion-varystrength} results for three values of
$\sigma_{ee}$ in steps of factors 2, namely for the standard value of
$\sigma_{ee}=6.5$ a$_0^2$ (see appendix \ref{sec:relaxtime}) used in
most of the paper, and for 13 $a^2_0$ (twice that cross section), and
for 3.25 a$_0^2$ (half of it). The comparison is done for the
one-plasmon excitation regime $E^*_0=2.7$ eV.  The effects of varying
$\sigma_{ee}$ are large. From the entropy signal (panel a), we can
read off that the global relaxation time shrinks here almost inversely
proportional to the cross section, from 100 fs over 50 fs down to
about 25 fs. One can spot the same trend in the attenuation of the
dipole signal in panel c.  Correspondingly, we see an increasing
suppression of ionization with increasing cross section. In this case,
however, the changes are more moderate, far slower than
proportional. The same holds for the anisotropy (panel b). Reduction
of $\beta_2$ increases with $\sigma_{ee}$, but rather slowly.

It is to be noted that the excitation of $E^*_0=2.7$ eV chosen here is
the most sensitive case for variation of $\sigma_{ee}$ around the
standard choice. Very little happens, of course, deep in the linear
regime where dissipation is negligible anyway. Somewhat less
sensitivity is also seen for heftier excitations (not shown here).

\subsubsection{Trends with system size}

\begin{figure}
\centerline{\includegraphics[width=0.9\linewidth]{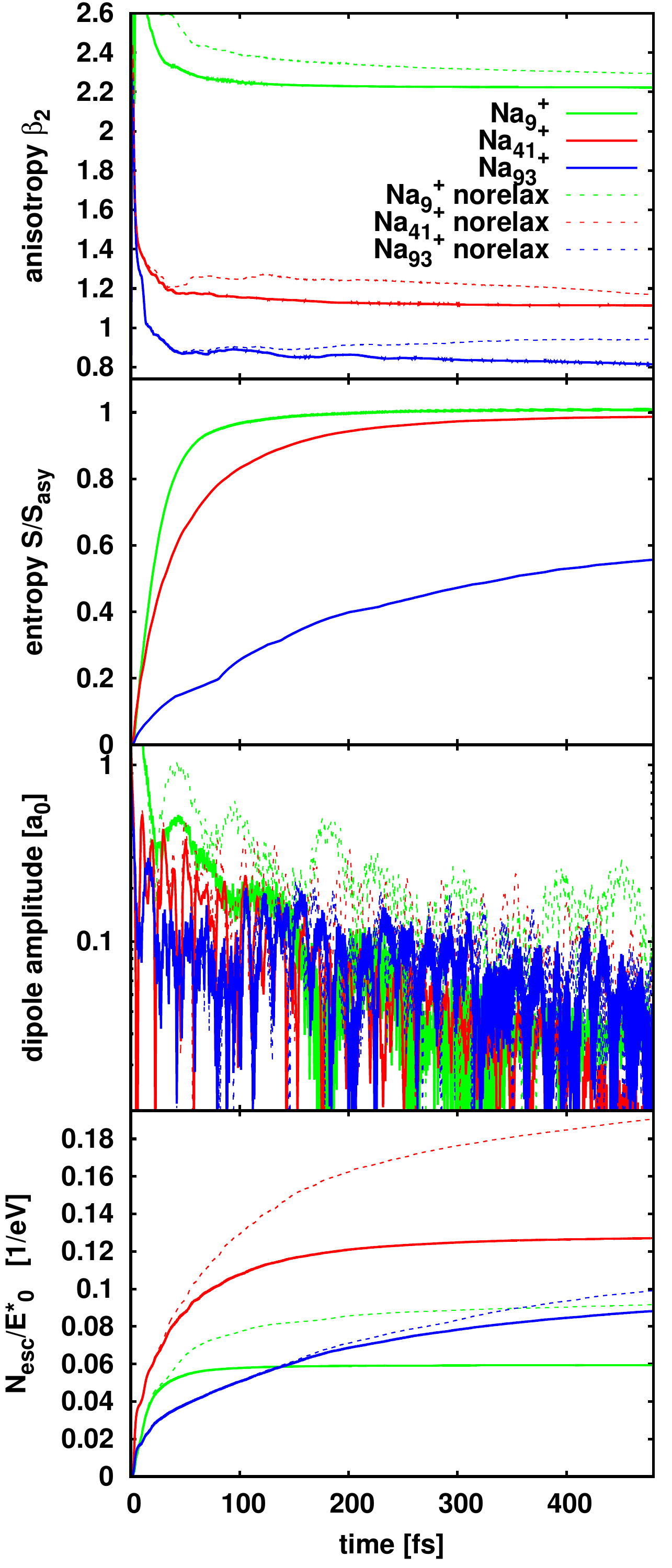}}
\caption{\label{fig:NaXXp-b20} 
 (Color online) Same as Fig.~\ref{fig:Na40ion-varyboost13}, but for varied 
cluster size comparing Na$_{9}^+$,  Na$_{41}^+$, and Na$_{93}^+$.
Boost energy is in all cases  $E^*_0=2.7$ eV.
}
\end{figure}
The next question is how dissipative effects depend on charge state
and system size. Figure \ref{fig:NaXXp-b20} shows results for three
cluster cations of different size, Na$_{9}^+$, Na$_{41}^+$, and
Na$_{93}^+$, all with the same excitation $E^*_0=2.7$ eV in the
one-plasmon regime. Comparison of Na$_{41}^+$ \PGR{here in Figure
  \ref{fig:NaXXp-b20} with the case of $E_0^*=$ 2.7 eV for Na$_{40}$
  in Figure \ref{fig:Na40ion-varyboost13}} shows that one more charge
changes very little in all respects \PGR{(emission, dipole amplitude,
  relaxation time, anisotropy)}.  However, system size makes a huge
difference. The smaller system Na$_{9}^+$ has much stronger
dissipative effects (shorter relaxation time, more suppression of
ionization) while the heavier cluster Na$_{93}^+$ shows very small
relaxation.  A quick glance at recipe (\ref{eq:relaxtime}) for the
relaxation rate explains this.  The rate is proportional to
$E^*_\mathrm{intr}/N$. Large electron number $N$ thus reduces the rate
while low $N$ enhances it. The effect is obvious: the plasmon energy
depends only weakly on system size \cite{Rei92a} but the thermal
effects are related to the energy per particle. The latter quantity
shrinks with increasing size. As a consequence, a one-plasmon
excitation remains safely in the linear regime for heavy clusters
while small clusters experience more thermal effects as, e.g.,
relaxation.

\subsection{Laser excitation}
\label{sec:laser}

Having explored the basic features of RTA for
in terms of the simple boost excitation, we
now present a quick first exploration of laser excitations.
The laser field is described as a classical electro-magnetic wave
handled in the limit of long wavelengths. This amounts to add to TDLDA
a time-dependent external dipole field
\begin{subequations}
\label{eq:Elaser}
\begin{eqnarray}
  U_\mathrm{ext}(\mathbf{r},t) 
  &=& 
  e^2\mathbf{r}\cdot\mathbf{e}_\mathrm{z} 
  \, E_0 \, f(t) \, \sin(\omega_\mathrm{las}t) 
  \quad,
\\
  f(t)
  &=&
  \sin^2\left(\pi\frac{t}{T_\mathrm{pulse}}\right)
  \theta(t)\theta(T_\mathrm{pulse}-t)
  \quad.
\end{eqnarray}
\end{subequations}
The laser features therein are: the (linear) polarization
$\mathbf{e}_\mathrm{z}$ along the symmetry axis, the peak field
strength $E_0$ related to laser intensity as $I_0\propto E_0^2$,
photon frequency $\omega_\mathrm{las}$, and pulse length
$T_\mathrm{pulse}$. Actually we use $T_\mathrm{pulse}=96$ fs
corresponding to a Full-Width at Half Maximum (FWHM)  of 48 fs for field
strength.

In order to track the energy balance in the process, we use here two
energetic observables: first  the energy absorbed from the laser field
\begin{equation}
 E_\mathrm{exc}
 =
 \int_0^t dt'\int d^3r\, \varrho(\mathbf{r},t')
 \partial_{t'}U_\mathrm{ext}(\mathbf{r},t')
\end{equation}
where $\varrho$ is the usual electron density, and second, the
intrinsic kinetic energy $E^*_\mathrm{intr}$ (Eq. (\ref{eq:Eintr})), see appendix \ref{app:eintr}.

\begin{figure}
\centerline{\includegraphics[width=0.9\linewidth]{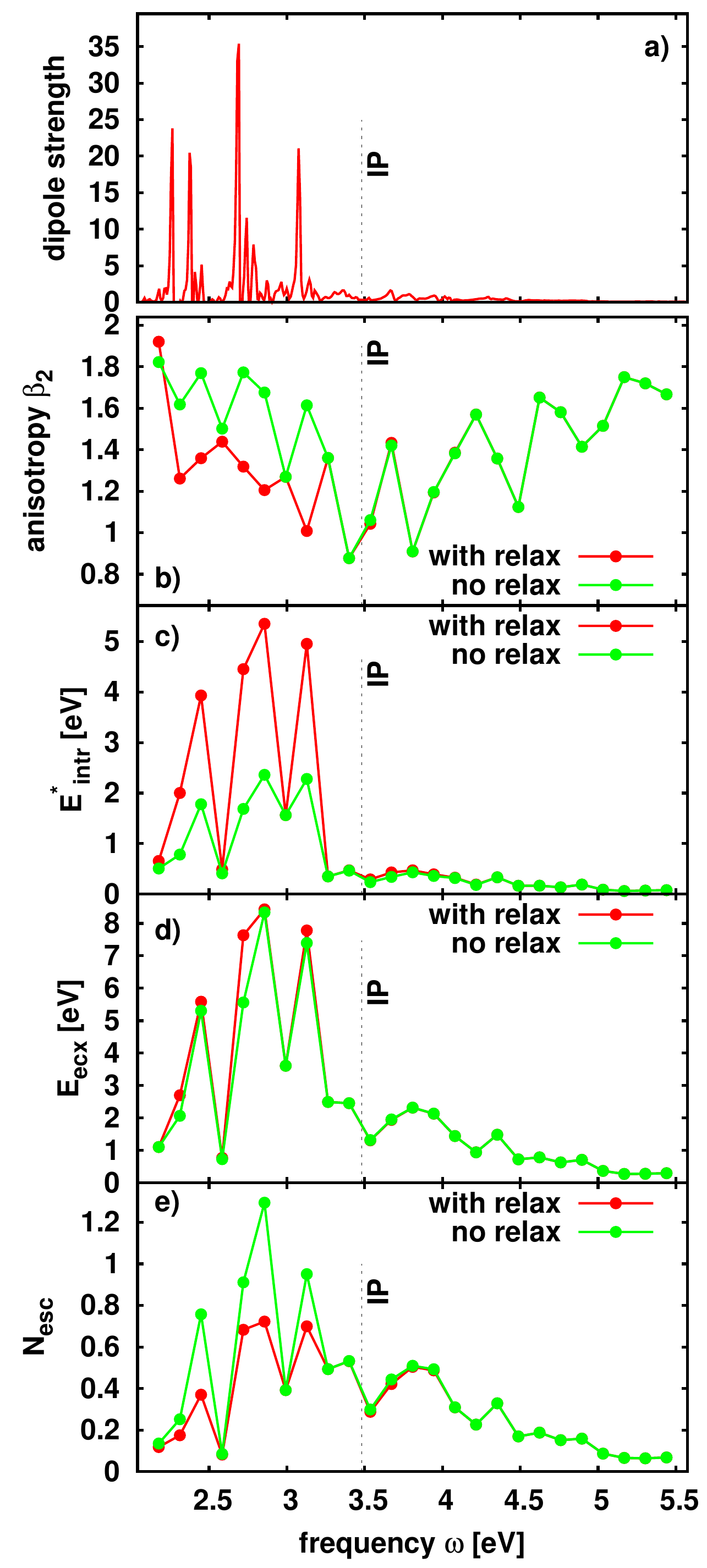}}
\caption{\label{fig:Na40-oscan_e0011_ion} 
 (Color online) Asymptotic values of four
  key observables after laser excitation with a laser pulse of
  intensity $I=10^{10}$W/cm$^2$, pulse with sin$^2$ profile of
  duration $T_\mathrm{pulse}=96$ fs (thus FWHM=48 fs), and varying
  frequency (drawn along $x$ axis).  Test case is Na$_{40}$ with ionic
  background in CAPS. The observables are: e) ionization
  $N_\mathrm{esc}$, d) excitation energy deposited by the laser
  $E_\mathrm{exc}$, c) intrinsic (thermal) kinetic energy
  $E^*_\mathrm{intr}$ and b) anisotropy $\beta_2$ of the angular
  distribution of emitted electrons, and entropy. Compared are results
  from RTA (using the standard scattering cross section 6.5 a$_0^2$) and
  TDLDA without relaxation. The uppermost panel a) shows for
  comparison the spectral distribution of dipole strength computed
  from an instantaneous boost (see section \ref{sec:optresp}).  }
\end{figure}
Figure \ref{fig:Na40-oscan_e0011_ion} shows results for a laser
intensity $I=10^{10}$W/cm$^2$. Calculations were done up to time 250
fs for a set of laser frequencies. Shown in panels b--e are the
asymptotic values of the observables which are practically reached at
this final time. Panel a shows for comparison the dipole strength
distribution obtained from spectral analysis after small boost (linear
regime).  The frequency dependence of ionization (panel e) maps
roughly the dipole strength distribution (compare with panel a and
mind that the present scan of $\omega_\mathrm{las}$ has low
resolution). Relaxation suppresses ionization considerably at the
peaks of the distribution, i.e. for resonant excitation. Practically
no dissipative effects are seen in the off-resonant minima between the
peaks and in the region above IP. Absence of dissipation is related to
an emission process faster than the relaxation time. This is obvious
for the energies above IP. These are direct emission processes which
run for Na clusters at a time scale of few fs \cite{Cal00,Rei04aB}.
Very instructive is the different behavior for on- and off-resonant
processes below IP. Resonant excitations are known to oscillate for a
longer amount of time \cite{Rei99a} which, in turn, gives dissipation
long time to interfere.  Off-resonant excitations, even if they
involve multi-photon processes, are confined to short times. The
photons involved have to cooperate instantaneously.

The intrinsic kinetic energy $E^*_\mathrm{intr}$ shown in panel c of
Figure \ref{fig:Na40-oscan_e0011_ion} complements the information from
ionization. No differences between TDLDA and RTA are seen for the
off-resonant processes while resonant processes shift a larger part of
the excitation energy to $E^*_\mathrm{intr}$, that part being thus
lost for direct ionization. It is interesting to note that the total
excitation energy $E_\mathrm{exc}$ absorbed from the laser (panel d)
is practically the same with and without relaxation term. It is the
balance between intrinsic excitation (thermalization) and direct
ionization which is regulated by the relaxation term. As expected,
anisotropy $\beta_2$ (panel b) is visibly lowered for all resonant
processes where dissipation is at work. \PGR{A similar effect had been
  see in previous study using VUU \cite{Gig03a}}.

\begin{figure}
\centerline{\includegraphics[width=0.9\linewidth]{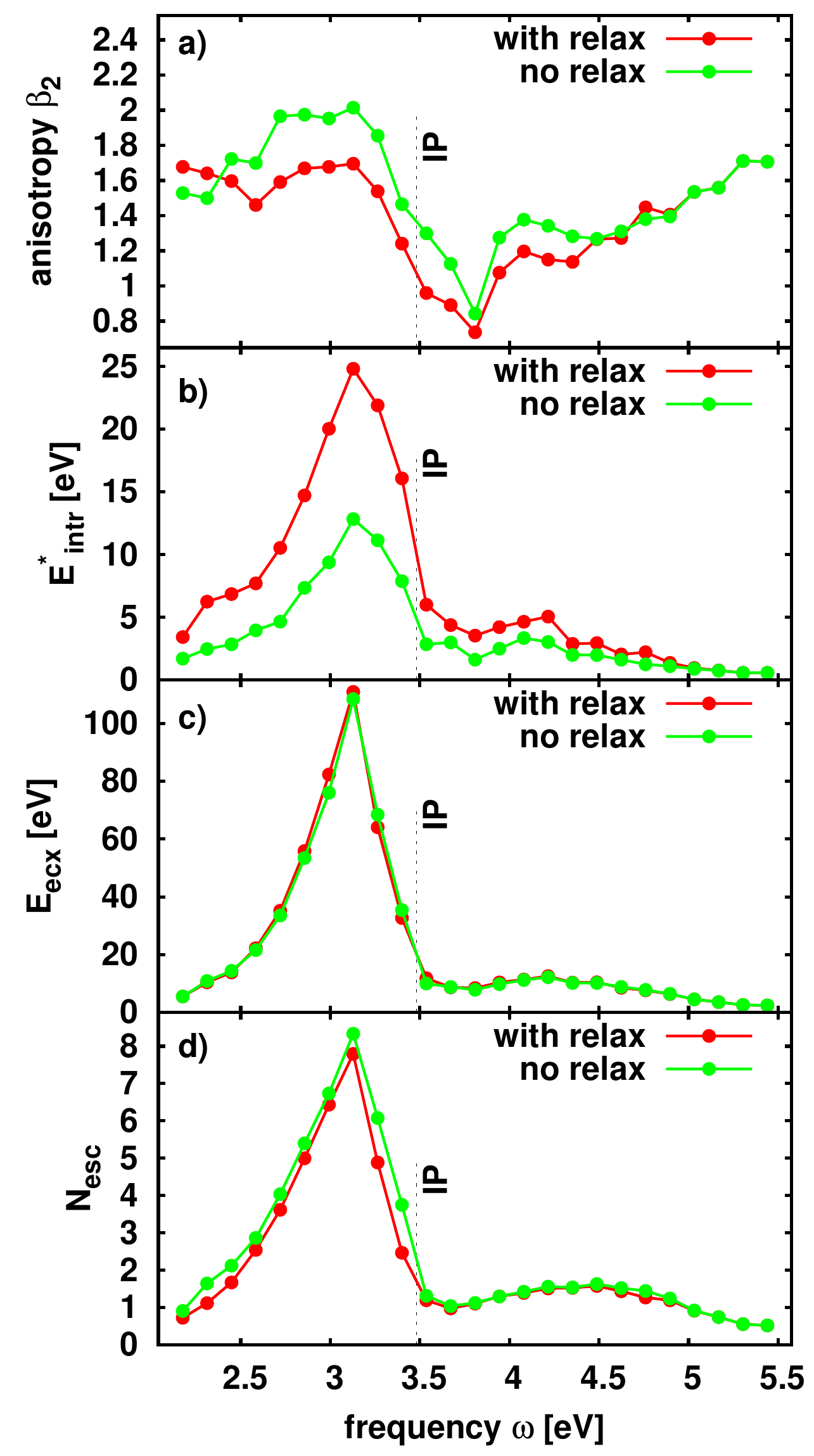}}
\caption{\label{fig:Na40-oscan_e0034_ion} 
 (Color online) Asymptotic values of four
  key observables after laser excitation with a laser pulse of
  intensity $I=10^{11}$W/cm$^2$, pulse with sin$^2$ profile of
  duration $T_\mathrm{pulse}=96$ fs (thus FWHM=48 fs), and varying
  frequency (drawn along $x$ axis).  Test case is Na$_{40}$ with ionic
  background in CAPS. The observables are: d) ionization
  $N_\mathrm{esc}$, c) excitation energy $E_\mathrm{exc}$ absorbed
  from the laser, b) intrinsic (thermal) kinetic energy
  $E^*_\mathrm{intr}$, a) anisotropy $\beta_2$ of the angular
  distribution of emitted electrons. 
  Compared are results from calculations with RTA
  (using the standard scattering cross section 6.5 a$_0^2$) and
  without. }
\end{figure}
Figure \ref{fig:Na40-oscan_e0034_ion} shows results for a higher laser
intensity $I=10^{11}$W/cm$^2$. The pattern are much different from the
previous case. Ionization (panel d) shows one broad resonance
peak. This happens already for TDLDA because ionization acts here as a 
strong dissipation mechanism \cite{Ull98a}. In fact, ionization is
always fast in this regime of violent excitations and thus always the
dominant mechanism. Little is left to do for collisional relaxation.
The energy absorption (panel c) is huge and, of course, shows no
difference with and without relaxation term. The effect of dissipation
is seen more clearly in the intrinsic energy (panel b). The ratio
between the full calculation and pure TDLDA amounts to about a  factor
two. Mind, however, that this is a large difference on a small
quantity. In any case, only a small fraction of the total absorbed
energy (panel c) is moved into intrinsic excitation and 80\% or 90\%
are eaten up for ionization. The $E^*_\mathrm{intr}$ shows another
feature which seems surprising at first glance: a difference between
RTA calculation and TDLDA persists in  some region above the IP. This has
a simple explanation. The strong ionization enhances the IP in the
course of the dynamics. Thus the system sees in the average a larger
IP than in the ground state and this shifts the regime of purely
direct processes to higher laser intensity. The fact that somewhat more
of the absorbed energy is moved into intrinsic excitation is also
reflected by the anisotropy (panel a) which is reduced as compared to
TDLDA up the point where  $E^*_\mathrm{intr}$ differs.

After all, these two short examples indicate that the interplay of
laser excitation and collisional relaxation carries a world of
interesting effects waiting to be uncovered. This calls for further
studies.

\section{Conclusions}
\label{sec:conc}

In this paper, we have proposed a practical way to include collisional
correlations in finite fermion systems at a quantum mechanical
level. The issue is crucial for energetic processes in many systems
from nuclei to clusters and molecules.  Our approach is inspired from
the semi-classical picture, but
remains strictly quantum mechanical. It relies on a Relaxation Time
Approximation (RTA) of the quantum collision term. The key ingredients
of our RTA are the instantaneous equilibrium density matrix
$\hat{\rho}_\mathrm{eq}(t)$ and the instantaneous relaxation time
$\tau_\mathrm{relax}(t)$. The scheme is applicable to any finite
fermion system. It turns out to be efficient and robust in various
dynamical scenarios.

As typical application we have investigated simple metal clusters
subject to a possibly strong electromagnetic perturbation (collision
with by-passing ion or laser irradiation). Inclusion of dissipation
through RTA provides the expected behaviors: enhanced damping of
oscillations, reduced ionization, more energy transfer to intrinsic
degrees of freedom (electronic thermalization), and more isotropic
emission of electrons. The effects strongly depend on excitation
conditions and strengths (boost amplitude, laser frequency or
intensity) which is very plausible. For the given collision rates used
in our RTA, we find a strong competition between collisional
relaxation and damping through direct electron emission whose outcome
depends sensitively on the actual dynamical conditions. RTA becomes
particularly beneficial for resonant laser excitations. In this case,
pure TDLDA is plagued by long lasting oscillations of the system which
are unphysical. RTA yields a realistic attenuation of the dipole
signal.

The present RTA provides a valuable extension of mean field theories
such as the LDA version of DFT for energetic dynamical scenarios
requiring a proper account of dissipation. Although RTA is restricted
to situations which can be modeled by one global (but instantaneous)
relaxation rate it certainly provides a valuable step in the right
direction. It surely deserves further exploration end extension to
other observables as, e.g., the widely explored photo-electron
spectra. Work along that line is in progress.

\section*{Acknowledgments}
We thank Thomas Fennel for helpful discussions.
We thank the RRZE (Regionales Rechenzentrum Erlangen) for supplying
the computing resources for the studies presented here. We also thank
Institut Universitaire de France, ITN network CORINF and french ANR
Muses program for financial support.

\appendix

\section{Solution of the RTA equations}
\label{sec:solveRTA}

\subsection{Interlaced mean-field and RTA time steps}

For completeness we
remind the RTA equation (\ref{eq:EoMbasic})
\begin{eqnarray*}
  \partial_t\hat{\rho}
  &=&
  -\mathrm{i}\big[\hat{h},\hat{\rho}\big]
  -
  \frac{1}{\tau_\mathrm{relax}}
  \left(\hat{\rho}-\hat{\rho}_\mathrm{eq}[\hat{\rho}]\right)
  \quad,
\end{eqnarray*}
Its r.h.s. contains first the
term (\ref{eq:KSwf}) driving mean-field propagation and second the
dissipative term. Mean-field propagation covers all s.p. oscillations
and runs at a much faster pace than relaxation. We thus treat the two
contributions at different time scales. The dissipative term is
evaluated in time steps of $\Delta t$ (thus at discrete times
$t_n=n\Delta t$), while the mean-field propagation
(\ref{eq:KSwf}\ref{eq:KSpropag}) is resolved on a finer mesh $\delta
t$. To make the dependencies more transparent, we express the
instantaneous equilibrium density more generally as
$\hat{\rho}_\mathrm{eq}[\hat{\rho}]$, i.e. as functional of given
density operator $\hat{\rho}$ which is communicated through the local
density, current, and energy of $\hat{\rho}$.  

Formally, we express the higher resolution for the mean-field
propagation in terms of the evolution operator (\ref{eq:KSpropag})
and use that to transform to the interaction picture.  During the step
from $t_n=n\Delta t$ to $t_{n+1} = (n+1) \Delta t$ we thus map
\begin{subequations}
\begin{eqnarray}
   \tilde{\rho}(t)
   &=&
   \hat{U}^{-1}(t,t_n)\hat{\rho}(t)\hat{U}(t,t_n)
   \;,
\label{eq:transfor}\\
   \hat{\rho}(t)
   &=&
   \hat{U}(t,t_n)\tilde{\rho}(t)\hat{U}^{-1}(t,t_n)
   \;,
\label{eq:transback}
\end{eqnarray}
\end{subequations}
which turns Eq. (\ref{eq:EoMbasic}) into
\begin{eqnarray*}
  \partial_t\tilde{\rho}
  &=&
  -
  \frac{1}{\tau_\mathrm{relax}}
  \left(\tilde{\rho}-\tilde{\rho}_\mathrm{eq}[\hat{\rho}]\right)
  \quad.
\end{eqnarray*}
Integrating time over the interval $[t_n,t_{n+1}]$ yields
\begin{eqnarray*}
  \tilde{\rho}(t_{n+1})
  &=&
  \tilde{\rho}(t_n)
  -
\int_{t_n}^{t_{n+1}}\!\! dt'\,
  \frac{
     \tilde{\rho}(t')
     -
     \tilde{\rho}_\mathrm{eq}[\hat{\rho}(t')]
    }{\tau_\mathrm{relax}}
\nonumber\\
  &\approx&
  \tilde{\rho}(t_n)
  -
  \frac{\Delta t}{\tau_\mathrm{relax}}
  \left[
     \tilde{\rho}(t_{n+1})
     -
     \tilde{\rho}_\mathrm{eq}[\hat{\rho}(t_{n+1})]
  \right]
\end{eqnarray*}
where the last step represents  a perturbative evaluation of the
dissipation in the time interval $[t_n,t_{n+1}]$.  
Using Eq. (\ref{eq:transback}), we transform back into the Schr\"odinger
picture to
\begin{subequations}
\begin{eqnarray}
  \hat{\rho}(t_{n+1})
  &=&
  \hat{\rho}_\mathrm{mf}
  -
  \frac{\Delta t}{\tau_\mathrm{relax}}
  \left[
  \hat{\rho}_\mathrm{mf}
     -
     \hat{\rho}_\mathrm{eq}[\hat{\rho}_\mathrm{mf}]
  \right]
  \;,
\label{eq:relaxstep}\\
  \hat{\rho}_\mathrm{mf}
  &=&
  \hat{U}(t_{n+1},t_n)\hat{\rho}(t_n)\hat{U}^{-1}(t_{n+1},t_n)
\end{eqnarray}
\end{subequations}
with all terms now expressed at time $t_{n+1}$.  This equation
delivers the stepping scheme. We first perform a mean-field
propagation in standard manner from $t_n$ to $t_{n+1}$.  This yields
the propagated $\hat{\rho}_\mathrm{mf}$. We then compute the
corresponding
$\hat{\rho}_\mathrm{eq}[\hat{\rho}_\mathrm{mf}]\equiv
\hat{\rho}_\mathrm{eq}[\varrho_\mathrm{mf},\mathbf{j}_\mathrm{mf},E_\mathrm{mf}]$, 
and
use that finally to compose the relaxation step (\ref{eq:relaxstep}).

\subsection{Estimate of the relaxation time}
\label{sec:relaxtime}

We refer to a semi-classical estimate of relaxation rates developed in
\cite{Ber78a} for homogeneous nuclear matter. Rescaling that from the
nuclear spin-isospin degeneracy factor $g=4$ to the electronic spin
degeneracy $g=2$, we obtain as starting point
\begin{equation}
  \frac{\hbar}{\tau_\mathrm{relax}}
  =
  3.95\frac{\hbar^2}{m}\sigma_{ee}k_\mathrm{F}\rho_0
  \frac{T^2}{\varepsilon_\mathrm{F}^2}
\end{equation}
where $m$ is the electron mass, $k_\mathrm{F}$ the Fermi momentum,
$\rho_0$ the matter density, $\varepsilon_\mathrm{F}$ the Fermi energy,
and $\sigma_{ee}$ the effective in-medium cross section for
electron-electron collisions. Using the standard relations
for a degenerate electron gas, 
$r_s={1.92}/{k_\mathrm{F}}$, $\varepsilon_\mathrm{F}={3.68}/{r_s^2}$,
and $\rho_0={3}/{(4\pi r_s^3)}$  (see \cite{Mar10aB}), we obtain
\begin{equation}
  \frac{\hbar}{\tau_\mathrm{relax}}
  =
  0.133\frac{\hbar^2}{m}\sigma_{ee}{T^2}
  \quad.
\end{equation}
It is preferable to express the rate in term of the intrinsic thermal
excitation energy \cite{Ber78a,Mar10aB}
\begin{equation}
  \frac{{E}^*_\mathrm{intr}}{N}
  =
  \frac{\pi^2 T^2}{4\varepsilon_\mathrm{F}}
\label{eq:Eth}
\end{equation}
where $N$ is the actual particle number. A word of caution is
necessary here. We are working in a system with absorbing boundary
conditions to account for ionization. This means that the number of
particles is a time dependent quantity. The value $N$ is the one
characterizing the system at a given instant and thus time
dependent.  We resolve the above expression to $T$ and express again
$\varepsilon_\mathrm{F}$ through $r_s$. This yields finally
Eq. (\ref{eq:relaxtime}), i.e.
${\hbar}/{\tau_\mathrm{relax}}={0.40}{{E}^*_\mathrm{intr}}{\sigma_{ee}}/(N{r_s^2})$.
The intrinsic excitation energy ${E}^*_\mathrm{intr}$ is a dynamical
observable which has to be determined anew at each time step, for
details see Appendix \ref{app:eintr}.

The Wigner-Seitz radius $r_s$ is a crucial parameter in the estimate
of the relaxation time. It characterizes the average electron density
and it is naturally given in the case of the jellium model. For a
cluster with detailed ionic background, there are two ways to deduce
$r_s$ from the given cluster. One can take either the electronic
diffraction radius $R_\mathrm{el,diffr}$ (box equivalent radius) as
defined in \cite{Fri82} and identify
\begin{subequations}
\begin{equation}
  r_s
  =
  R_\mathrm{el,diffr}N^{-1/3}
\end{equation}
or one can take the ionic r.m.s. radius $r_\mathrm{ion}$ which is
often simpler to evaluate. The ionic structure has practically no
surface zone and thus the ionic diffraction radius is
$R_\mathrm{el,diffr}=\sqrt{3/5}r_\mathrm{ion}$. We know that the
electron distribution in metals follows closely the ionic background
due to the strong, attractive Coulomb interaction.  This allows to
identify alternatively
\begin{equation}
  r_s
  =
  \sqrt{\frac{3}{5}}\,r_\mathrm{ion}N^{-1/3}
  \quad.
\label{eq:rs-def-ion}
\end{equation}
\end{subequations}
Both definitions yield in practice very similar results.  We use the
form (\ref{eq:rs-def-ion}) which is simpler to handle, particularly in
case where the ionic structure is frozen. In practice, it is
sufficiently well described by a constant $r_s=3.7$ a$_0$.

It remains to determine the effective cross section $\sigma_{ee}$.  We
use here the careful evaluation of \cite{Koe08a,Koe12a}.  They compute
electron screening for homogeneous electron matter in Thomas-Fermi
approximation, compute from that the scattering cross-section and
apply a Pauli correction of factor 1/2. This yields
$\sigma_\mathrm{ee}=6.5$ a$_0^2$ for the case of Na clusters at
$r_s\approx 3.7$ a$_0$. It is this value which was used as reference
throughout this paper.

\section{Density Constrained Mean-Field (DCMF)}
\label{sec:hdenscurrE}

A key task is to determine the instantaneous equilibrium density-operator
$\hat{\rho}_\mathrm{eq}[\varrho,\mathbf{j},E]$ in the RTA equation
(\ref{eq:EoMbasic}). It is the mean-field state of minimum energy
under the constraints of given local density $\varrho$, current
$\mathbf{j}$, and energy $E$.  A scheme for
density constrained mean-field (DCMF) calculations was developed in
\cite{Cus85a}.  We use it here in a version extended to account also
for the constraint on current
$\mathbf{j}(\mathbf{r})$.
The density constrained mean-field Hamiltonian then reads
\begin{eqnarray}
  \hat{h}_\mathrm{dens.co.}[\varrho]
  &=&
  \hat{h}_{mf}[\varrho]
  -
  \int d^3r\,\lambda_\varrho(\mathbf{r})\hat{\varrho}(\mathbf{r})
\nonumber\\
  &&\qquad
  -
  \int d^3r\,\mbox{\boldmath$\lambda_j$}(\mathbf{r})\hat{\mathbf{j}}(\mathbf{r})
\end{eqnarray}
where $\hat{h}_{mf}[\varrho]$ is the standard mean-field Hamiltonian
for the given local density $\varrho(\mathbf{r})$,
$\hat{\varrho}(\mathbf{r})$ is the operator of local density,
$\hat{\mathbf{j}}(\mathbf{r})$ the operator of local current, while
$\lambda_\varrho$ and {\boldmath$\lambda_j$} stand for the associated Lagrange
parameters. These are determined iteratively such that the solution of
the corresponding Kohn-Sham equations yields the wanted density
$\varrho(\mathbf{r})$ and current $\mathbf{j}(\mathbf{r})$ \cite{Cus85a}.

As a further constraint, we want to adjust the equilibrium state to
given energy $E_\mathrm{goal}=E_\mathrm{mf}=E[\hat{\rho}]$ and
particle number $N_\mathrm{goal}$. $N_\mathrm{goal}$ represents the
actual number of particles in the system, also denoted $N$ in the core
of the text of the paper. Its computation is simple. It is just
given by the integral over the computational box of the density
$\varrho({\bf r})$ at a given RTA time step $t_n$. 
The adjustment of $E_\mathrm{goal}$ and $N_\mathrm{goal}$ is
achieved by considering a thermalized mean-field state
$\hat{\rho}_\mathrm{eq}$ of the form Eq. (\ref{eq:rhodiag}) where each
s.p. state $\phi_\alpha$ is augmented with an occupation weight
\begin{subequations}
\begin{eqnarray}
  W_\alpha^\mathrm{(eq)}
  &=&
  \frac{1}{1+\exp{\left((\varepsilon_{\alpha}-\mu)/T\right)}}
  \;,
\label{eq:Fermi}
\\
  (\mu,T)
  &\leftrightarrow&
  \mbox{tr}\{\hat{\rho}_\mathrm{eq}\}
  =
  N_\mathrm{goal}
  \;,\;
  E[\hat{\rho}_\mathrm{eq}]
  =
  E_\mathrm{goal}
  \;.
\end{eqnarray}
\end{subequations}
Temperature $T$ and chemical potential $\mu$ are to be adjusted such
that the wanted total particle number $N_\mathrm{goal}$ and energy
$E_\mathrm{goal}$ is reproduced. To simplify the computations, the
occupation numbers are adjusted to the Fermi form (\ref{eq:Fermi})
once before DCMF and once after the DCMF iterations. The final
reoccupation after DCMF slightly spoils the reproduction of density
and current. But the deviations remain small along all time
propagation.

\section{The intrinsic excitation energy}
\label{app:eintr}

DCMF is also used to compute the intrinsic excitation energy
$E^*_\mathrm{intr}$ which
is, in fact, the intrinsic kinetic energy. The total kinetic energy is the
difference between the actual energy $E$ and the minimal energy at
given local density $\varrho(\mathbf{r})$ which is the energy of the
(stationary) DCMF state for fixed $\varrho(\mathbf{r})$,
$\mathbf{j}=0$ and $T=0$.  Part of the total kinetic energy goes into
the kinetic energy of the collective flow represented by
$\mathbf{j}(\mathbf{r})$. The other part is the intrinsic kinetic
energy ${E}^*_\mathrm{intr}$. We evaluate it by computing the DCMF
state for given $\varrho(\mathbf{r})$, $\mathbf{j}(\mathbf{r})\neq 0$,
but now for $T=0$ to find the minimum energy 
$E_\mathrm{DCMF}[\varrho(\mathbf{r}),\mathbf{j}(\mathbf{r}),T\!=\!0]$
under the given
constraints. The difference 
between total energy and this energy is then the intrinsic
kinetic energy
\begin{equation}
  {E}^*_\mathrm{intr}
  =
  E
  -
  E_\mathrm{DCMF}[\varrho(\mathbf{r}),\mathbf{j}(\mathbf{r}),T\!=\!0]
  \quad.
\label{eq:Eintr}
\end{equation}
This fully quantum-mechanical definition is used for estimating the
relaxation time in Eq. (\ref{eq:relaxtime}) which is a crucial
parameter in the propagation. For analyzing purposes (see section
\ref{sec:laser}), we can use the less expensive semi-classical
Thomas-Fermi approximation for $E_\mathrm{kin,DCMF}$ as done in
previous works \cite{Cal00,Rei04aB}.

\section{Mixing of two one-body density matrices}
\label{sec:mix}

The dissipative step delivers a density matrix (\ref{eq:relaxstep})
which is a mix
\begin{eqnarray*}
  \hat{\rho}(t_{n+1})
  &=&
  \left(1-\eta\right)\hat{\rho}_\mathrm{mf}
  +
  \eta\hat{\rho}_\mathrm{eq}
  \;,\;
  \eta
  =
  \frac{\Delta t}{\tau_\mathrm{relax}}
  \;,
\\
  \hat{\rho}_\mathrm{mf}
  &=&
  \sum_{\alpha}|\phi_\alpha^\mathrm{(mf)}\rangle
  W_\alpha\langle\phi_\alpha^\mathrm{(mf)}|
  \;,
\\
  \hat{\rho}_\mathrm{eq}
  &=&
  \sum_{\alpha}|\phi'_\alpha\rangle
  W'_\alpha\langle\phi'_\alpha|
  \;,
\end{eqnarray*}
where the 
$|\phi_\alpha^\mathrm{(mf)}\rangle=\hat{U}(t_{n+1},t_n)|\phi_\alpha(t_n)\rangle$ 
constitute the basis of
TDLDA-propagated states while the $|\phi'_\alpha\rangle$ are new
states from DCMF which also delivers new occupancies $W'_\alpha$.  We
expand the composed state $\hat{\rho}(t_{n+1})$ with respect to the
set $\{|\phi_\alpha^\mathrm{(mf)}\rangle\}$ because it represents the majority
contribution. It reads
\begin{eqnarray*} 
  &&\hat{\rho}(t_{n+1})
  =
  \sum_{\alpha\beta}|\phi_\alpha^\mathrm{(mf)}\rangle
  \rho_{\alpha\beta}\langle\phi_\beta^\mathrm{(mf)}|
  \;,
\\
  &&
  \rho_{\alpha\beta} 
  =
  (1\!-\!\eta)\delta_{\alpha\beta}W_\alpha
  + 
  \eta\sum_{\gamma}\langle\phi_\alpha^\mathrm{(mf)}|\phi'_{\gamma}\rangle
  W'_{\gamma}\langle\phi'_{\gamma}|\phi_\beta^\mathrm{(mf)}\rangle 
  \;.
\end{eqnarray*}
For the further propagation, we want to use again the diagonal
representation (\ref{eq:rhodiag}). To this end, we diagonalize
$\rho_{\alpha\beta}$ which finally delivers the new set
$|\phi_\alpha(t_{n+1})\rangle$ and $W_\alpha(t_{n+1})$ which will
remain constant over the next RTA step from $t_{n+1}$ to $t_{n+2}$.

\section{Iterative correction of particle number and energy}
\label{sec:corriter}

After mixing (\ref{eq:relaxstep}), one may end up with a slight
deviation from the wanted energy $E_\mathrm{goal}$ although we have
taken care that both entries as such meet this energy (see section
\ref{sec:hdenscurrE}).  But minimal energy violations may accumulate
during long propagation.  The task is thus to correct for a possibly
occuring, small energy error
\begin{equation}
  \delta E
  =
  E(\hat{\rho}_\mathrm{mix})-E_\mathrm{goal}
\end{equation}
where $E(\hat{\rho}_\mathrm{mix})$ is the actual total energy after
the mixing step.  We need a scheme to correct the total energy with as
little modification as possible. To this end, we
may employ an iterative strategy. We avoid direct use of the
equilibrium distribution and express a corrective step in terms of the
given occupation numbers $W_\alpha$. To this end, we take over from
the equilibrium distribution the crucial property
\begin{equation}
  \delta W_\alpha^\mathrm{(equi)}
  =
  -\left(
   \delta T\frac{\varepsilon_\alpha-\mu}{T^2}
   +
   \delta\mu\frac{1}{T}
  \right)
  W_\alpha^\mathrm{(equi)}\big(1-W_\alpha^\mathrm{(equi)}\big)
  \quad.
\label{eq:change-occ}
\end{equation}
We take this as motivation to postulate for a general change
\begin{equation}
  \delta W_\alpha
  =
  \left(\delta_1\varepsilon_\alpha+\delta_0\right)
  W_\alpha(1-W_\alpha)
  \quad.
\label{eq:change-occ2}
\end{equation}

The $\delta_i$ are tuned to desired change in particle number $\delta N$
and in energy $\delta E$ (where $\delta E$ is determined
from the energy loss in the relaxation step). The conditions to fulfill are 
\begin{equation*}
  \sum_\alpha\delta W_\alpha
  =
  0
  \quad,\quad
  \sum_\alpha\varepsilon_\alpha \delta W_\alpha
  =
  \delta E
  \quad.
\end{equation*}
The energy conservation is formulated in terms of the s.p. energies.
This is valid in the linear regime because of
\begin{equation*}
  E(\hat{\rho}+\delta\hat{\rho})
  =
  E(\hat{\rho})+\mathrm{tr}\{\hat{h}\delta\hat{\rho}\}
  =
  E(\hat{\rho})+
  \sum_\alpha \delta W_\alpha\varepsilon_\alpha
  \quad.
\end{equation*}
Thus we want to fulfill
\begin{subequations}
\begin{eqnarray}
  \sum_\alpha\delta W_\alpha
  =
  \delta N
  &=&
  \delta_0\overline{1}+\delta_1\overline{\varepsilon}
  \quad,
\\
  \sum_\alpha\delta W_\alpha\varepsilon_\alpha
  =
  \delta E
  &=&
  \delta_0\overline{\varepsilon}+\delta_1\overline{\varepsilon^2}
  \quad,
\end{eqnarray}
\end{subequations}
with $\overline{A}=\sum_\alpha w_\alpha(1-w_\alpha)A_\alpha$.
This linear system is resolved by
\begin{subequations}
\begin{eqnarray}
  \delta_0
  &=&
  \frac{\overline{\varepsilon^2}\delta N-\overline{\varepsilon}\delta E}
       {\overline{\varepsilon^2}\overline{1}-(\overline{\varepsilon})^2}
  \quad,
\\
  \delta_1
  &=&
  \frac{-\overline{\varepsilon}\delta N+\overline{1}\delta E}
       {\overline{\varepsilon^2}\overline{1}-(\overline{\varepsilon})^2}
  \quad.
\end{eqnarray}
\end{subequations}
Thus we obtain the $\delta_i$ and
subsequently the $\delta W_\alpha$ according to
eq. (\ref{eq:change-occ2}). Finally, it has to be checked whether the
readjusted density matrix
$\hat{\rho}_\mathrm{mix,cor}=\sum_\alpha|\phi_\alpha\rangle(W_\alpha+\delta{W}_\alpha)\langle\phi_\alpha|$
fulfills $E(\hat{\rho}_\mathrm{mix,cor})=E_\mathrm{goal}$.  In case,
the energy matching is not good enough, the above stepping has to be
repeated.

%

\bibliographystyle{apsrev}
\bibliography{relaxtime}

\end{document}